\documentclass[pra,twocolumn,showpacs,floatfix,amsfonts]{revtex4}
\usepackage{bm}
\usepackage{graphicx}
\usepackage{here}
\usepackage{latexsym}
\usepackage{amsmath}
\usepackage{amssymb}
\usepackage{dcolumn}

\begin{document}

\newcommand{\x}{{\bf x}}
\renewcommand{\k}{{\bf k}}
\newcommand{\E}{{\bf E}}
\newcommand{\e}{{\bf e}}
\def\mn#1{\marginpar{*\footnotesize \footnotesize #1}{*}}
\def\isum{{\sum\!\!\!\!\!\!\!\!\int}}
\def\bfzero{{\bf 0}}
\newcommand{\ket}[1]{\, | #1 \rangle}
\newcommand{\be}{\begin{equation}}
\newcommand{\ee}{\end{equation}}
\newcommand{\bea}{\begin{eqnarray}}
\newcommand{\eea}{\end{eqnarray}}
\newcommand{\bdm}{\begin{displaymath}}
\newcommand{\edm}{\end{displaymath}}
\newcommand{\bfd}{{\bf d}}
\newcommand{\bfk}{{\bf k}}
\def\hF{{\bar F}}
\newcommand{\hbE}{{\hat {\bf E}}}
\newcommand{\hbP}{{\hat {\bf P}}}
\newcommand{\hpsi}{{\hat {\psi}}}
\renewcommand{\k}{{\bf k}}
\renewcommand{\r}{{\bf r}}
\newcommand{\bfp}{{\bf p}}
\newcommand{\bfq}{{\bf q}}
\newcommand{\bfE}{{\bf E}}
\newcommand{\bfP}{{\bf P}}
\newcommand{\ha}{{\hat a}}
\newcommand{\bfr}{{\bf x}}
\newcommand{\bfz}{{\bf 0}}
\newcommand{\cl}{{\cal l}}
\newcommand{\cE}{{\cal E}}
\newcommand{\tG}{{\widetilde G}}
\newcommand{\veps}{\varepsilon}
\newcommand{\non}{\nonumber \\}
\newcommand{\mGamma}{\bar \Gamma}

\def\gsym{\gamma}

\def\pder#1#2{\frac{\partial #1}{\partial #2}}
\def\lsim{\raise0.3ex\hbox{$<$\kern-0.75em\raise-1.1ex\hbox{$\sim$}}}

\title{Spatial effects in superradiant Rayleigh scattering from Bose-Einstein 
  condensates}
\author{O.\ Zobay and Georgios M.\ Nikolopoulos}
\affiliation{Institut f\"ur Angewandte Physik, Technische Universit\"at
Darmstadt, 64289 Darmstadt, Germany}
\date{\today}
\begin{abstract}
We present a detailed theoretical analysis of superradiant Rayleigh 
scattering from atomic Bose-Einstein condensates. A thorough investigation   
of the spatially resolved time-evolution of optical and matter-wave fields 
is performed in the framework of the semiclassical 
Maxwell-Schr\"odinger equations. 
Our theory is not only able to explain many of the known experimental 
observations, e.g., the behavior of the atomic side-mode distributions, but also provides further detailed insights into the coupled dynamics of optical and matter-wave fields. 
To work out the significance of propagation effects, we compare our results to other theoretical models in which these effects are neglected.
\end{abstract}
\maketitle

\section{Introduction}
\label{sec:Int}

The beautiful recent experiments of Refs.\ \cite{InoChiSta99,SchTorBoy03} have opened up the possibility of studying superradiance within the context of Bose-Einstein condensates of ultracold atomic gases. In a typical experimental setup, a cigar-shaped condensate is exposed to a far-off resonant laser pulse travelling in a direction perpendicular to the long condensate axis [see Fig.\ \ref{fig_BECsuper}(a)]. Condensate atoms can then undergo Rayleigh scattering thereby experiencing a recoil kick. The moving atoms together with the condensate at rest form matter-wave gratings from which further laser photons are scattered. This in turn produces additional recoiling atoms, causing the amplitudes of the matter-wave gratings and the scattered light fields to grow rapidly in a self-amplifying process. Due to phase-matching effects induced by the elongated condensate shape, the fastest growth is experienced by those gratings for which the scattered photons travel parallel to the condensate axis in the two so-called optical ``endfire modes." After an initial start-up and mode competition process \cite{MooMey99}, the ensuing superradiant light emission is concentrated into the endfire modes, and the recoiling atoms have well-defined momenta thus forming two first-order atomic ``side modes" [Fig.\ \ref{fig_BECsuper}(a)].

Comparing this process to ``conventional" superradiance in an inverted atomic medium \cite{Dic54,GroHar82}, we see that the role of the electronically excited atoms is now taken by the combination of the condensate and the applied laser pulse, spontaneous photon emission is replaced by Rayleigh scattering, and de-excited atoms correspond to atoms in the momentum side modes \cite{SchTorBoy03}. Another important difference to the conventional process, which stops after each atom has undergone a single transition, is the possibility of repeated or modified scattering cycles which lead to the population of multiple atomic side modes. These processes occur naturally if the applied laser pulse is of sufficient duration or strength. On the one hand, an atom in a side mode can again scatter a laser photon, thereby transferring to a higher-order side mode. Alternatively, however, atoms may also interact with endfire mode photons and scatter them back into the laser field \cite{SchTorBoy03}. This leads to the production of atoms moving backwards with respect to the direction of the applied laser field. Since the latter process requires to overcome an energy mismatch, it only occurs for sufficiently intense laser pulses. In this way, one is led to distinguish between the weak- and the strong-pulse regimes of superradiance. The behavior of the system strongly differs in these two regimes. In particular, one important difference concerns the atomic side-mode population patterns which can be investigated through time-of-flight imaging. In the weak-pulse regime, where forward scattering is prevalent, distributions take a characteristic fan shape, whereas in the strong-pulse regime, an X-like pattern is observed.

Basic features of the atomic momentum distributions were theoretically discussed by a number of authors (see, e.g., \cite{MusYou00,PuZhaMey03,BenBen04,VasEfiTri04,RobPioBon04}) without, however, giving a comprehensive description of the experimental observations. Reference \cite{RobPioBon04}, for example, examined a model in which two regimes of either pure forward scattering or combined backward and forward scattering could be distinguished depending on the external parameters. However, this study only considered a strictly one-dimensional system so that a direct comparison to the experiments was not possible.

A more comprehensive investigation of the atomic side-mode distributions was recently presented in our work \cite{ZobNik05}. In Ref.\ \cite{ZobNik05}, the typical X- and fan-shape patterns along with some of their characteristic properties were reproduced and explained in terms of the underlying coupled dynamics of optical and matter-wave fields. Our approach is built upon two main concepts: First of all, we use a semiclassical description for the field dynamics. Following earlier studies of conventional superradiance \cite{GroHar82,HaaKinSch79a,HaaKinSch79b}, we can expect that the semiclassical approach is valid as soon as the numbers of atoms and photons in the atomic side and optical endfire modes, respectively, become large compared to one. It is this macroscopic regime of superradiance that we are focussing on throughout this work. A fully quantized model, on the other hand, such as the one used by Meystre and co-workers \cite{MooMey99,PuZhaMey03}, allows to investigate the initial startup of the superradiant process, but is not easily extended to the study of long-time dynamics. The second main feature of our approach is the inclusion of spatial propagation effects. This aspect has been neglected in almost all previous quantal as well as semiclassical treatments of Bose-Einstein condensate (BEC) superradiance (see, e.g., \cite{MooMey99,MusYou00,PuZhaMey03,BenBen04,VasEfiTri04,RobPioBon04}). References \cite{AveTri04,BonPioRob05} examine some spatial effects in the BEC-light interaction, but do not provide a detailed comparison to the experimental observations of Refs.\ \cite{InoChiSta99,SchTorBoy03}. The results of Ref.\ \cite{ZobNik05}, however, show that the inclusion of spatial effects is crucial for a full understanding of BEC superradiance. In fact, our results also allowed us to resolve the controversy between Refs.\ \cite{SchTorBoy03} and \cite{PuZhaMey03} regarding the origin of the spatial asymmetry between forward- and backward-moving atomic side modes observed in the strong-pulse regime of superradiance.

This successful explanation of several key observations of the superradiance experiments \cite{InoChiSta99,SchTorBoy03} suggests that the model of Ref.\ \cite{ZobNik05} provides an adequate description of the system dynamics in the semiclassical regime. The purpose of the present paper is therefore to extend our previous work by providing an in-depth examination of this model and, in this way, to obtain further detailed insights into the system dynamics. We also continue to stress the significance of spatial propagation effects by explicitly comparing our results to a related mean-field model in which these effects are neglected (see, e.g., \cite{RobPioBon04}). 

The paper is organized as follows. In Sec.\ II, we present our theoretical framework which is based on the semiclassical Maxwell-Schr\"odinger equations and the use of the slowly-varying-envelope approximation. We also indicate how one may include photon scattering within or between the optical endfire modes. Section III analytically discusses the short-time gain regime. In particular, it is shown that due to the propagation effects, the growth of the atomic side-mode populations and optical endfire-mode intensities is slower than exponential. In Sec.\ IV, we discuss the strong-pulse regime and explain in detail how the characteristic observations of X-shape side-mode patterns and the spatial asymmetry come about. We also exhibit explicitly the failure of spatially independent models to account for these effects. In Sec.\ V, the weak-pulse regime is investigated. Besides discussing the atomic side-mode distribution patterns, we also use our model to study characteristic spatial effects in the coupled dynamics of optical and matter-wave fields. It will turn out that these effects are significantly more involved than in the strong-pulse regime. The paper concludes with a short summary and outlook in Sec.\ VI.

\section{Maxwell-Schr\"odinger Equations}
\label{sec:Model}

\begin{figure}[t]
\centerline{\includegraphics[width=7.cm]{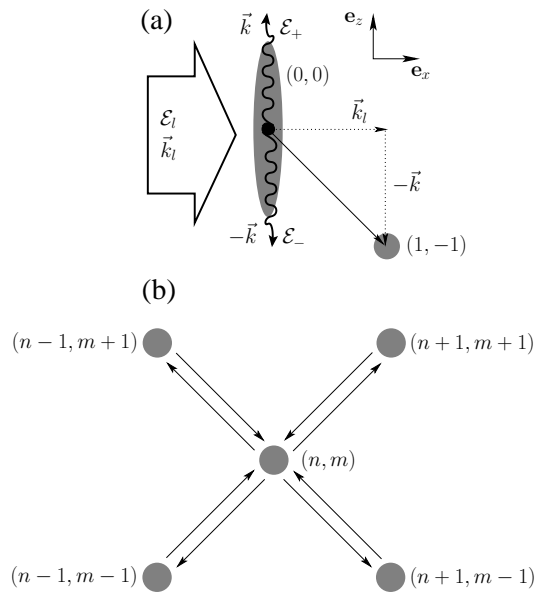}}
\caption{ Schematic representation of BEC 
superradiance. (a) A cigar-shaped BEC 
of length $L$ (filled ellipsoid) is exposed to a linearly polarized laser 
pulse with amplitude $\cE_l(t)$ and wave vector ${\bf k}_l$. 
Stimulated Rayleigh scattering of laser photons off the condensate produces recoiling atoms in side modes with well-defined momenta. The side mode 
$(0,0)$ refers to the BEC at rest whereas only one of the two first-order  side modes, i.e., $(1,-1)$, is shown. The scattered photons 
propagate mainly along the condensate axis in the endfire modes $\cE_\pm$.
(b) Population transfer to and from the side mode $(n,m)$ due to interaction with the laser pulse as implied by Eq.\ (\protect\ref{psi_scal}). 
\label{fig_BECsuper}}
\end{figure}

Our model involves an elongated condensate of length $L$ 
oriented along the $z$ axis. 
The BEC consists of $N$ two-level atoms which are coupled via the 
electric-dipole interaction to a linearly polarized pump laser pulse  
$\cE_l(t){\bf e}_y (e^{i(k_lx-\omega_l t)}+c.c.)/2$
with $\omega_l = ck_l$, traveling in the $x$ direction 
[see Fig. \ref{fig_BECsuper}(a)]. 
The laser is considered to be far off resonant from the atomic transition 
to the excited electronic state $\ket e$. 

As discussed in the Introduction, in this system the coherent nature of the 
BEC leads to strong correlations between successive Rayleigh scattering 
events and to collective 
superradiant behaviour \cite{InoChiSta99,SchTorBoy03,MooMey99}. 
Moreover, as a result of the cigar shape of the 
condensate, the gain is largest when the scattered photons leave the 
condensate travelling up and down its long axis in the so called endfire 
modes $\cE_\pm(\x,t)\e_y e^{-i(\omega t\mp kz)}$, where $\cE_\pm(\x,t)$ are  
the envelope functions of the modes. 
As a consequence, the recoiling atoms have well-defined momenta and appear in distinct atomic side modes.  In the side mode $(n,m)$, 
atoms possess momentum $\hbar(nk_l\e_x + mk\e_z)$, while their kinetic 
energy is given by $\hbar\omega_{n,m} = \hbar^2(n^2k_l^2+m^2k^2)/2M$. 
In this notation, the ``side mode" $(0,0)$ describes the condensate at rest. 
The wave vectors $\pm k {\bf e}_z$ of the scattered photons are fixed by energy conservation 
for the transitions between the side modes $(0,0)$ and $(1,\pm 1)$ which 
initiate the process, i.e., $\hbar ck_l = \hbar ck +\hbar \omega_{11}$.
Given that $\omega_{n,m}\ll\omega_l$, we approximately have $k\approx k_l$ 
and $k_l-k \ll k,k_l$. Thus, the side-mode frequency is approximately 
given by $\omega_{n,m} \approx (n^2+m^2)\omega_r$, where 
$\omega_r = \hbar k_l^2/2M$ is the recoil frequency.  

Maxwell-Schr\"odinger equations of motion for the coupled matter-wave and 
electromagnetic fields offer a very general theoretical 
framework for the description of problems pertaining to the interaction 
of ultracold atoms with light \cite{CohDupGry89,ZhaWal94,Mey01}. For the 
problem at hand, 
after adiabatically eliminating the excited atomic state $\ket e$, the coupled 
Maxwell-Schr\"odinger equations for the {\em mean-field} macroscopic 
atomic wave function $\psi(\x,t)$ and the positive- and negative-frequency 
components $\bfE^{(\pm)}(\x,t)$ of the classical electric field read 
\cite{GroHar82,CohDupGry89,ZhaWal94}
\bea\label{psi_ad}
i\hbar\pder{}t\psi &=& -\frac{\hbar^2}{2M}\Delta\psi+\frac{(\bfd\cdot\bfE^{(-)})(\bfd\cdot\bfE^{(+)})} {\hbar\delta}\psi,\\
\label{E_field}
\pder{^2 \bfE^{(\pm)}}{t^2} &=& c^2 \Delta \bfE^{(\pm)}-\frac 1{\veps_0} \pder{^2 \bfP^{(\pm)}}{t^2}
\eea
with $\delta$ the detuning of the laser from the electronic transition, $\bfd$ 
the atomic dipole moment and $M$ the atomic mass. The polarization is given by 
\bea
\bfP^{(+)}(\x,t)= -\bfd |\psi(\x,t)|^2 
\frac{\bfd\cdot\bfE^{(+)}(\x,t)}{\hbar\delta}, 
\eea
with $\bfP^{(-)}= \bfP^{(+)\,*}$.
Note that in Eq.\ (\ref{psi_ad}) we neglect the external trapping potential 
and interactions between atoms, since they do not play a significant role on the time scales of 
the process under consideration. However, it would be straightforward to 
include them in the model. 

In order to solve Eqs.\ (\ref{psi_ad}) and (\ref{E_field}) under the 
slowly-varying-envelope approximation (SVEA), we decompose the fields as
\bea\label{svea_psi}
\psi(\x,t) &=& \sum_{(n,m)} \frac{\psi_{nm}(z,t)}{\sqrt A} e^{-i(\omega_{n,m}t-nk_lx - mkz)},\\ \label{svea_ep}
\E^{(+)}(\x,t) &=& \cE_l\e_y e^{-i(\omega_l t-k_lx)}/2 +\cE_+(z,t)\e_y e^{-i(\omega t-kz)}\non
&& + \cE_-(z,t)\e_y e^{-i(\omega t+kz)},
\eea
$\E^{(-)}(\x,t)=\E^{(+)\, *}(\x,t)$, with $\omega=kc$ and $A$ the average condensate cross section perpendicular to the $z$ axis. 
The electric field in Eq.\ (\ref{svea_ep}) contains the impinging laser 
pulse together with the two optical endfire modes that are produced by 
collective Rayleigh scattering.  
The atoms in the side mode $(n,m)$ are characterized by a slowly varying 
spatial envelope $\psi_{nm}(z,t)$. The summation in 
Eq.\ (\ref{svea_psi}) can be restricted to terms $(n,m)$ with $m+n$ even, as Eq.\ (\ref{psi_scal}) below will show.

Note that in the ansatz (\ref{svea_psi}) and (\ref{svea_ep}) we disregard the 
dependence of the envelope functions $\psi_{nm}$ and $\cE_\pm$ on the 
transverse directions $x$ and $y$. 
For the matter waves, this is certainly a good approximation since the 
radial degrees of freedom are tightly confined by the trap. 
Moreover, the use of the same approximation for the optical fields is 
justified by the fact that in the experiments the Fresnel number of the system is close 
to $1$ \cite{InoChiSta99,GroHar82}.

In order to introduce a concise notation, we rescale the optical and matter waves 
as  
\bea
\cE_{\pm,l} &\to&  e_{\pm,l}\sqrt{\frac{\hbar\omega k_l}{2\veps_0 A}};\quad 
\psi_{nm} \to \frac{\psi_{nm}\sqrt{k_l}}{\sqrt{A}}.
\eea
Then, using the ansatz (\ref{svea_psi}) and (\ref{svea_ep}) and introducing the 
dimensionless time $\tau=2\omega_{r}t$ and length $\xi=k_lz$, 
Eq.\  (\ref{psi_ad}) in the SVEA reads
\begin{widetext}
\bea
\label{psi_scal}
i\pder{\psi_{nm}(\xi,\tau)}{\tau} &=&  -\frac 12 \pder{^2 \psi_{nm}(\xi,\tau)}{\xi^2}-im\pder{\psi_{nm}(\xi,\tau)}{\xi}\non
&& +\kappa \Big[ e_+^*(\xi,\tau)\psi_{n-1,m+1}(\xi,\tau)e^{i(n-m-2)\tau}
+e_-^*(\xi,\tau)\psi_{n-1,m-1}(\xi,\tau) e^{i(n+m-2)\tau}\non
&& +e_+(\xi,\tau)\psi_{n+1,m-1}(\xi,\tau)e^{-i(n-m) \tau}+e_-(\xi,\tau)\psi_{n+1,m+1}(\xi,\tau) e^{-i(n+m)\tau}\Big]\non
&& +\lambda\Big[e_-^*(\xi,\tau)e_+(\xi,\tau)\psi_{n,m-2}(\xi,\tau) e^{2i(m-1)\tau}+e_+^*(\xi,\tau)e_-(\xi,\tau)\psi_{n,m+2}(\xi,\tau) e^{-2i(m+1)\tau}\Big]\non
&&+\lambda(|e_+(\xi,\tau)|^2+|e_-(\xi,\tau)|^2) \psi_{nm}(\xi,\tau),
\eea
\end{widetext}
with the coupling constants 
\begin{subequations}
\bea
\kappa &=& \frac{g}{2\omega_r}\sqrt{k_lL},\\
\lambda &=& \frac{\kappa}{e_0},
\eea
\end{subequations}
where
\bea
g = \frac{|\bfd|^2\cE_l}{2\hbar^2\delta} 
\sqrt{\frac{\hbar\omega_l}{2\veps_0 AL}}.
\eea 
The first term on the right-hand side of Eq. (\ref{psi_scal}) describes the 
quantum-mechanical dispersion of the envelope function, while the second one 
leads to a spatial translation with velocity $v_m = m\hbar k/M$. 
The other terms describe the interaction between the matter-wave and 
electromagnetic fields. This interaction leads to spatially 
dependent shifts and couplings of the momentum side mode $(n,m)$ to other 
modes. Let us discuss the underlying physical processes in more detail.

The terms involving the coupling constant $\kappa$ refer to photon exchange 
between one of the endfire modes and the laser beam [see Fig.\ \ref{fig_BECsuper}(b)].
In particular, through 
stimulated scattering, an atom in a side mode $(n,m)$ can absorb a laser 
photon and deposit it into one of the endfire modes. 
The accompanying recoil transfers the atom into one of the side modes 
$(n+1,m\pm 1)$. Alternatively, the atom may absorb an endfire-mode photon and 
emit it into the laser beam, thereby ending up in the side mode 
$(n-1,m\pm 1)$. This latter process is responsible for atomic backward 
scattering \cite{SchTorBoy03}. 

Another class of physical processes involves the photon exchange between 
the two side modes and is described by the terms with $e_\pm e_\mp^*$. 
In this case the accompanying recoil transfers the atom from $(n,m)$ to the side modes $(n,m\pm 2)$.    
Finally, the terms containing $|e_\pm(\xi,\tau)|^2$ describe the absorption 
of a photon from an endfire mode and its subsequent emission into the same 
mode. Hence, such processes give only rise to shifts which depend on 
both time and space, but do not introduce couplings to other side modes.
The corresponding term for the laser beam 
is related to a constant ac Stark shift and as such is not included in 
Eq. (\ref{psi_scal}). 

In the SVEA, the envelope functions $e_\pm$ of the endfire modes obey the 
equations 
\begin{widetext}
\bea
\label{Ep_scal}
\pder{e_+}\tau+\chi \pder{e_+}\xi &=&-i\sum_{(n,m)}\Big[\kappa e^{i(n-m) \tau} \psi_{nm}(\xi,\tau)\psi_{n+1,m-1}^*(\xi,\tau)\non
&&+\lambda e_-(\xi,\tau)e^{-2i(m-1)\tau} \psi_{nm}(\xi,\tau)\psi_{n,m-2}^*(\xi,\tau) +\lambda e_+(\xi,\tau)|\psi_{nm}(\xi,\tau)|^2\Big],\\
\label{Em_scal}
\pder{e_-}\tau-\chi \pder{e_-}\xi &=& -i\sum_{(n,m)}\Big[\kappa e^{i(n+m) \tau} \psi_{nm}(\xi,\tau)\psi_{n+1,m+1}^*(\xi,\tau)\non
&&+\lambda e_+(\xi,\tau)e^{2i(m+1)\tau} \psi_{nm}(\xi,\tau)\psi_{n,m+2}^*(\xi,\tau)+\lambda e_-(\xi,\tau)|\psi_{nm}(\xi,\tau)|^2\Big]
\eea
\end{widetext}
where
\bea
\chi &=& \frac{ck_l}{2\omega_r}.
\eea

Introducing the atomic natural linewidth 
$\Gamma_{a}=d^2\omega^3/(3\pi\veps_0\hbar c^3)$, the coupling constant $g$ 
can be expressed in terms of the Rayleigh scattering 
rate $R=d^2\cE_l^2\Gamma_{ a}/(4\hbar\delta^2)$ as
\bea
g = \sqrt{\frac{3\pi c^3 R}{2\omega^2 A L}}.
\eea
Moreover, the collective superradiant gain is given by 
\bea
G= \frac{g^2 N}{\gamma},
\eea
with the photon damping rate $\gamma= 2c/L$ (note slight differences from Ref.\ \cite{RobPioBon04} regarding the numerical prefactors in the definitions of $G$ and $\gamma$). 
Experimental observations clearly distinguish between two different regimes 
of parameters, the so-called strong- and weak-pulse regimes 
\cite{InoChiSta99,SchTorBoy03,MooMey99}.  
The former regime is characterized by a scattering rate comparable to the 
recoil frequency and $G\gg\omega_r$, while for the latter $R\ll\omega_r$ 
and $G\leq \omega_r$. 

For both of these regimes, one can verify that photon exchange 
between endfire modes can be disregarded in our model (at least for the 
time-scales of interest) \cite{InoChiSta99,SchTorBoy03}. 
Hence, Eq. (\ref{psi_scal}) may be simplified to
\begin{subequations}
\label{sdeq}
\bea\label{env_psi}
&&i\pder{\psi_{nm}(\xi,\tau)} \tau = -\frac 1{2} \pder{^2 \psi_{nm}}{\xi^2} 
-i m \pder{\psi_{nm}}\xi\\
&&+\kappa\big[e_+^*\psi_{n-1,m+1}e^{i(n-m-2)\tau} 
+e_-^*\psi_{n-1,m-1} e^{i(n+m-2)\tau}\nonumber\\ 
&&+e_+\psi_{n+1,m-1}e^{-i(n-m)\tau}+e_-\psi_{n+1,m+1} e^{-i(n+m)\tau}\big]. \nonumber
\eea
Since the interaction between the optical and the matter-wave fields is restricted to the condensate volume, any relevant retardations are expected 
to be of the order of $L/c\simeq 10^{-12}\,$s and can be neglected.  
Thus, formal integration of Eqs. (\ref{Ep_scal})-(\ref{Em_scal}) 
for $\lambda=0$ yields the following equations for the envelope 
functions $e_\pm$:
\bea\label{e_p}
e_+(\xi,\tau) &=& -i\frac{\kappa}\chi\int_{-\infty}^\xi d\xi' 
\sum_{(n,m)} e^{i(n -m)\tau}\non
&& \times \psi_{nm}(\xi',\tau)\psi_{n+1,m-1}^*(\xi',\tau),\\
\label{e_m}
e_-(\xi,\tau) &=& -i\frac{\kappa}\chi\int^{\infty}_\xi d\xi' 
\sum_{(n,m)} e^{i(n+m)\tau}\non
&& \times \psi_{nm}(\xi',\tau)\psi_{n+1,m+1}^*(\xi',\tau).
\eea
\end{subequations}
From these equations one may clearly see how the buildup of the endfire-mode 
fields at $(\xi,t)$ is driven by the coherences $\psi_{nm}\psi_{n+1,m\pm 1}^*$. 

As mentioned in the Introduction, previous theoretical studies of BEC 
superradiance have neglected the spatial dependence of optical and 
matter-wave fields. 
It is one of the main purposes of this paper to compare 
the model derived above to this simpler approach in order to assess the 
significance of propagation effects. 
Formally, we can obtain the equations of motion for the spatially 
independent model from Eqs.\ (\ref{sdeq}) by dropping any spatial 
dependence and setting $\psi_{nm}\to C_{nm}/\sqrt{k_l L}$ 
and $e_\pm\to b_\pm/\sqrt{k_l L}$. 
The amplitudes $C_{nm}(\tau)$ and $b_\pm(\tau)$ can be interpreted as 
amplitudes of the matter-wave and optical fields, respectively, and 
they obey the equations of motion
\begin{subequations}
\label{sieq}
\bea
\label{env_psi_si}
&&\!\!\!\!\!\!\frac{dC_{nm}(\tau)} {d\tau} = -i\frac{g}{2\omega_{\rm r}}\Big [b_+^*C_{n-1,m+1}e^{i(n-m-2)\tau}\non
&&  +b_-^*C_{n-1,m-1} e^{i(n+m-2)\tau}+b_+C_{n+1,m-1}e^{-i(n-m)\tau}\non
&&+b_-C_{n+1,m+1} e^{-i(n+m)\tau}\Big ],\\
\label{e_p_si}
&&\!\!\!\!\!\! b_+(\tau) = -i\frac{g}{\gamma}
\sum_{(n,m)} e^{i(n -m)\tau}C_{nm}(\tau)C_{n+1,m-1}^*(\tau),\\
\label{e_m_si}
&&\!\!\!\!\!\! b_-(\tau) = -i\frac{g}{\gamma}
\sum_{(n,m)} e^{i(n+m)\tau}C_{nm}(\tau)C_{n+1,m+1}^*(\tau).
\eea
\end{subequations}
A similar model, which distinguished only the momentum side modes in the $x$ diretion, was presented in \cite{RobPioBon04}.
It should be emphasized that one cannot consider $C_{nm}(\tau)$ 
and $b_\pm(\tau)$ as spatial averages of $\psi_{nm}(\xi,\tau)$ and 
$e_\pm(\xi,\tau)$. 
In fact, we will see that the predictions of the two models show 
significant differences. It should also be noted that the damping 
coefficient $\gamma$, although of the order of $c/L$, cannot unambiguously 
be ascribed a precise value.

In general, superradiant scattering is initiated by quantum-mechanical noise, i.e., spontaneous Rayleigh scattering from individual condensate atoms \cite{MooMey99}. Subsequent stimulated scattering and bosonic enhancement lead to rapid growth of the side-mode populations. The semiclassical model derived in this section describes the macroscopic stage of the superradiant process where the populations of the side modes are already large compared to one. 

Adapting the discussion of Refs.\ \cite{GroHar82,HaaKinSch79a,HaaKinSch79b} regarding ``conventional" superradiance, we can take the effects of the initial quantum-mechanical fluctuations into account by solving the semiclassical equations of motion with stochastic initial conditions (seeds) for the side modes $\psi_{1,\pm 1}(\xi,\tau=0)$. Since the noise is practically relevant for the initial population of those modes only, we can set $\psi_{nm}(\xi,\tau=0) = 0$ for all other side modes. The condensate $\psi_{00}$ is chosen to be in the macroscopic ground state at $t=0$. For any of these stochastic initial conditions, the solution of the semiclassical equations of motion corresponds to one possible realization of the experiment. By studying a large set of simulations with varying seed functions drawn from an appropriate distribution, one could obtain information on, e.g., averages and fluctuations of relevant experimental observables. In the present paper, however, we intend to focus on characteristic features of the macroscopic dynamics that are observed in individual experimental realizations, such as those reported in Refs.\ \cite{InoChiSta99,SchTorBoy03}.

To this end, we have numerically solved Eqs.\ (\ref{sdeq}) with fixed external parameters for a large set of different seed functions, and we find the characteristic qualitative features, e.g., the side-mode distribution patterns, to show up independent of the choice of the specific initial condition. This shows that these features are caused by the macroscopic dynamics and are not related to quantum fluctuation effects. In Sec.\ \ref{secIII} we will explain how this insensitivity to the initial conditions comes about.

\section{Early stage of superradiant scattering}
\label{secIII}
In this section, we investigate the early stages of the superradiant process in detail. More precisely, we wish to examine the regime where the populations of the first-order side modes still remain far below the number of atoms in the condensate, but are large compared to $1$, so that the semiclassical model is applicable. The undepleted-pump approximation for the condensate can then be invoked, which allows to derive analytical results for the spatial distributions of the side-modes and their total populations. With these results, we can (i) compare the growth of the side modes in the weak- and strong-pulse regimes, (ii) work out differences to the spatially-independent model introduced at the end of Sec.\ II, and (iii) explain why the time-evolution of the side modes is rather insensitive to the details of the initial seed function. The most interesting result of these studies is probably the observation that the propagation effects lead to a subexponential growth of the side-mode populations which is in contrast to the spatially independent model where an exponential growth is predicted.

In the startup regime described above, only the first-order side modes for forward- and backward-scattering become populated significantly, so that we can restrict Eqs.\ (\ref{sdeq}) to these modes and the condensate. Furthermore, the depletion of the condensate is assumed to be negligible so that we set $\psi_{00}(\xi,\tau)\approx\psi_{00}(\xi,0)$ (undepleted-pump approximation). The equations for the sets of modes $(\pm 1, \pm 1)$ and $(\pm 1, \mp 1)$ then decouple. Neglecting the effects of free propagation, which is well justified for short times, the time evolution of the side modes $(\pm 1, \mp 1)$, for example, is governed by
\begin{subequations} \label{startup}
\bea
i\frac{\partial \psi_{1,-1}(\xi,\tau)}{\partial \tau} &=& i\frac{\kappa^2}{\chi}\psi_{00}(\xi)\int_{-\infty}^\xi d\xi'\big[e^{2i\tau}\psi_{-1,1}^*(\xi',\tau) 
\nonumber\\ && \!\!\! \times
\psi_{00}(\xi')+\psi_{00}^*(\xi')\psi_{1,-1}(\xi',\tau)\big],\\
i\frac{\partial \psi_{-1,1}(\xi,\tau)}{\partial \tau} &=& -i\frac{\kappa^2}{\chi}\psi_{00}(\xi)\int_{-\infty}^\xi d\xi'\big[\psi_{-1,1}(\xi',\tau) \nonumber \\&&  \times
\left. \psi^*_{00}(\xi')+ e^{2i\tau}\psi_{00}(\xi')\psi_{1,-1}^*(\xi',\tau)\right].
\eea
\end{subequations}
Before discussing the solutions of Eqs.\ (\ref{startup}), let us briefly 
consider the corresponding startup regime in the spatially independent model. 
Setting $C_{00}(\tau) = \sqrt{N}$ and $C_{-1,1}(\tau=0) = 0$, 
one obtains from Eqs. (\ref{sieq})
\begin{subequations} \label{cpm-cmp}
\bea\label{cpm}
C_{1,-1}(\tau) &=& \frac{C_{1,-1}(0)}{4(i-\lambda_1)}\left[e^{\lambda_1 \tau}
\left(2\lambda_2-\Gamma\right)
-e^{\lambda_2 \tau}\left(2\lambda_1-\Gamma\right)\right],\non\\
\label{cmp}\widetilde C_{-1,1}(\tau) &=& \frac{C_{1,-1}(0)}{4(i-\lambda_1)} \Gamma \left(e^{\lambda_1 \tau} - e^{\lambda_2 \tau}\right)
\eea
\end{subequations}
with $\widetilde C_{-1,1}(\tau) = e^{2i\tau} C^*_{-1,1}(\tau)$, $\Gamma = G/\omega_r$, and the growth rates $\lambda_1 = i-\sqrt{-1-i\Gamma}$, $\lambda_2 = i+\sqrt{-1-i\Gamma}$. In the weak-pulse regime of $\Gamma \ll 1$, one has $C_{1,-1}(\tau)\approx C_{1,-1}(0) e^{\Gamma \tau/2}$ and $|C_{-1,1}/C_{1,-1}|^2\approx \Gamma^2/16$. For strong pulses with $\Gamma \gg 1$, on the other hand, one finds $C_{1,-1}(\tau)\approx C_{1,-1}(0)\sqrt{\Gamma/16}~
 e^{(1-i)\sqrt{\Gamma/2}\tau+i\pi/4}$
and $|C_{-1,1}/C_{1,-1}|\approx 1$ (see also \cite{RobPioBon04}). We thus see that the spatially independent model always predicts an exponential increase of the side-mode populations, but with growth rates varying between $\Gamma$ and $\sqrt{\Gamma}$.

To facilitate the comparison with these predictions, we examine the spatially dependent equations (\ref{startup}) for constant initial conditions $\psi_{1,-1}(\xi,0) = \psi_0$, $\psi_{-1,1}(\xi,0) = 0$ and a homogeneous condensate $\psi_{00}(\xi) = \sqrt{N/\Lambda}$, $0 \le \xi \le \Lambda$. Here, $\Lambda=k_lL$ denotes the scaled condensate length. As shown in the Appendix, Eqs.\ (\ref{startup}) can then be solved approximately by Laplace transform and subsequent inversion through the saddle-point method. In the weak-pulse limit, one obtains
\be\label{psi_weak}
\psi_{1,-1}(\xi,\tau) \approx \frac{\psi_0}{\sqrt{4\pi}\sqrt[4]{\Gamma\tau\xi/\Lambda}} \exp\left(2\sqrt{\Gamma\tau\xi/\Lambda}\right),
\ee
so that the side-mode population grows like
\be\label{N_weak}
N_{1,-1}(\tau)=\int_0^{\Lambda}d\xi |\psi_{1,-1}(\xi,\tau)|^2 \approx \frac{N_{1,-1}(0)}{8\pi\Gamma\tau}\exp\left(4\sqrt{\Gamma\tau}\right)
\ee
with the initial population $N_{1,-1}(0) = |\psi_0|^2 \Lambda$.

In the strong-pulse regime, we find
\bea\label{psi_strong}
\!\!\!\psi_{1,-1}(\xi,\tau) &\approx & \frac{\psi_0 e^{5\pi i/12}}{2^{2/3}\sqrt{6\pi}} \left(\frac{\Gamma\xi}\Lambda\right)^{1/6}\frac 1{\tau^{2/3}}\non
&& \!\!\times\exp\left(i\tau+3e^{-i\pi/6}(\Gamma\tau^2\xi/2\Lambda)^{1/3}\right).
\eea
The total side-mode population is approximately given by
\be\label{N_strong}
N_{1,-1}(\tau) \approx \frac{N_{1,-1}(0)}{12\pi\sqrt 3}\frac 1{\tau^2} \exp\left(\frac{3^{3/2}}{\sqrt[3] 8} \Gamma^{1/3} \tau^{2/3} \right).
\ee

In contrast to Eqs. (\ref{cpm-cmp}), equations (\ref{N_weak}) and (\ref{N_strong}) show that propagation effects lead to a subexponential growth in the side-mode populations: for weak pulses, $N_{1,-1}$ increases as $\exp(\sqrt{\tau})/\tau$, whereas for strong pulses, it grows like $\exp(\tau^{2/3})/\tau^2$. Moreover, from Eqs.\ (\ref{psi_weak}) and (\ref{psi_strong}), we see that the side modes also grow subexponentially {\it in space}, but again display different behaviors in the weak- and strong-pulse limits.

For a spatially varying condensate wave function, e.g., a BEC ground state in a trap, it becomes more difficult to derive analytical results for the side modes.
However, the numerical solution of Eqs.\ (\ref{startup}) with the condensate having the Thomas-Fermi 
shape yields side-mode populations that are very close to the ones 
for a homogeneous condensate with the same atom number. 
We thus expect laws similar to Eqs.\ (\ref{N_weak}) and (\ref{N_strong}) to 
apply in this case as well. 

The study of Eqs.\ (\ref{startup}) also allows us to understand why, after an initial transient, the spatial shape of the side-mode wave functions becomes rather insensitive to the details of the initial seed function. To this end, we consider the Fourier decomposition of an arbitrary initial seed
\be\label{decom}
\psi_{1,-1}(\xi,0) = \sum_{n=-\infty}^\infty c_n \exp(i 2\pi n \xi /\Lambda).
\ee
Since Eqs.\ (\ref{startup}) are linear, $\psi_{1,-1}(\xi,\tau)$ is obtained as the linear superposition $\sum_n c_n \psi^{(n)}_{1,-1}(\xi,\tau)$, where $\psi^{(n)}_{1,-1}(\xi,0)=\exp(i 2\pi n \xi /\Lambda)$. Considering the weak-pulse regime for concreteness, the Laplace transform method shows that, after some time, the wave functions $\psi^{(n)}_{1,-1}$ behave like
\be\label{k_shape}
\psi^{(n)}_{1,-1}(\xi,\tau)\approx  \frac{ \exp\left(2\sqrt{\Gamma\tau\xi/\Lambda}\right)} {\sqrt{4\pi}\left(\sqrt[4]{\Gamma\tau\xi/\Lambda}-i{\displaystyle \frac{2\pi n}\Lambda} \xi/\sqrt[4]{\Gamma\tau\xi/\Lambda}\right)}.  
\ee
From this result, we can draw two conclusions. (i) All wave functions $\psi^{(n)}_{1,-1}$ acquire more or less the same shape, i.e., their moduli grow subexponentially in space, whereas their phases vary only slowly. We can thus expect their superposition to qualitatively show the same behavior. (ii) The contributions of higher-order Fourier components are suppressed by a factor of $|n|$, approximately. In fact, for growing $|n|$, we find numerically that it takes increasingly more time to reach the form $(\ref{k_shape})$, and before this time is reached, the suppression is even stronger. These observations explain the insensitivity of the side-mode wave functions to the details of the initial seed.

\section{Strong-pulse regime}

In Secs.\ IV and V, we will use the theoretical model developed in Sec.\ II to present a thorough discussion of the coupled dynamics of optical and matter waves in the strong- and weak-pulse regimes of superradiance. For the numerical calculations, we focus on the 
experimental data of Refs.\ \cite{InoChiSta99,SchTorBoy03}. 
In particular, we consider a $^{87}$Rb BEC of $N=2\times 10^6$ atoms, 
with length $L=200\mu$m and cross-section diameter $15\mu$m. 
The condensate is in the Thomas-Fermi regime, so that we can model its wave 
function as $\psi_{00}(z) = \sqrt{n(z)}$ with 
$n(z) = C[(L/2)^2-z^2]\Theta(L/2-|z|)$, $C=3N/4L^3$.  
As discussed in Secs.\ II and III, the results are not significantly 
influenced by the shape of the seed function. 
Hence, for the sake of simplicity, throughout our simulations the 
first-order atomic side modes are seeded according to 
$\psi_{1,\pm 1}(z,0) = \psi_{00}(z)/\sqrt N$, which corresponds to one 
delocalized atom in each of the modes. 
The applied laser pulse is modeled as rectangular 
lasting from $t=0$ up to $t=t_f$. Equations (\ref{sdeq}) are then propagated 
using a split-step algorithm \cite{BaoJakMar03}. The grid $(n,m)$ for the 
side-mode orders is chosen sufficiently large, so that the population of 
the highest-order side modes remains negligible at all times. The chosen 
values for the external parameters, such as pulse strength and duration, 
are comparable to those used in the experiments 
\cite{InoChiSta99,SchTorBoy03}.

Let us first consider the regime of strong laser pulses which 
is characterized by Rayleigh scattering rates $R$ comparable to  $\omega_r$. 
In this regime of parameters, experiments with short laser pulses have shown forward- as well backward-recoiling atoms, forming a very characteristic X-shape pattern. 
Moreover, there is a noticeable spatial asymmetry between forward and 
backward peaks. According to Ref.\ \cite{SchTorBoy03}, these observations 
suggest a purely optical picture of superradiance in which atoms are 
diffracted from the optical grating formed by the endfire and pump modes. 
The intensity of the endfire modes, and thus the optical grating, are assumed to be peaked 
at the edges of the condensate. Our model has enabled us to reproduce the experimental observations and to verify this conjecture.  

\begin{figure}
\centerline{\includegraphics[width=7.cm]{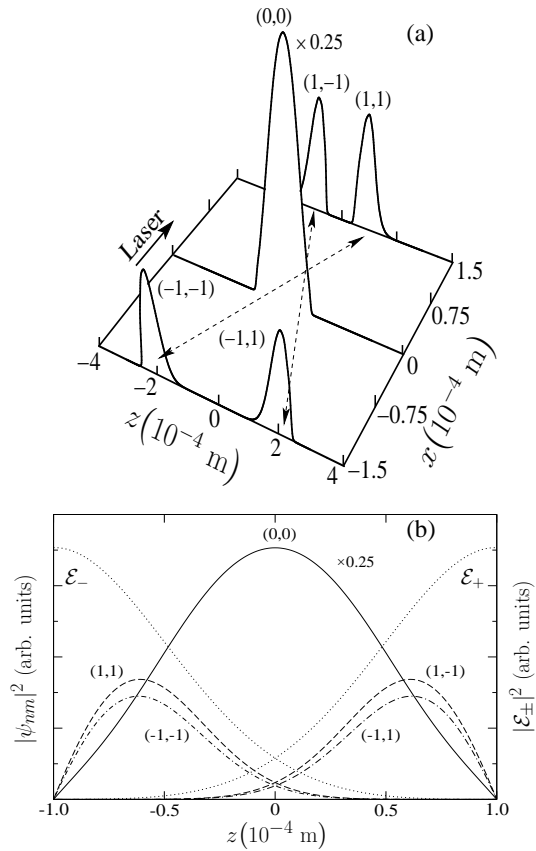}}
\caption{Spatially dependent model.
(a) Spatial distribution of the first-order forward $(1,\pm 1)$ and 
backward $(-1,\pm 1)$ atomic side modes, after applying a laser pulse of 
duration $t_f=8.5$ $\mu$s and strength $g=3\times 10^6$ s$^{-1}$ to the condensate 
followed by a free propagation for a time $t_p=25\,$ms.
(b) Spatial distributions of the atomic side modes and the optical 
endfire modes ${\cal E}_\pm$ at time $t_f$. 
For the sake of 
illustration, in both cases the population of the BEC $(0,0)$   
has been multiplied by $0.25$.
\label{fig1}}
\end{figure}

\subsection{Side-mode patterns}
In Fig.\ \ref{fig1}(a), we display a snapshot of the atomic spatial 
distribution 
after applying a strong laser pulse to the condensate followed by a  
free propagation for a time $t_p\gg t_f$. Since we work with a one-dimensional 
model, we calculate the displacement $\Delta x$ between the condensate and 
the first-order side modes in the $x$ direction as $\Delta x=v_r t_p$ with the 
recoil velocity $v_r=\hbar k/M=5.9\times 10^{-3}$ m/s. Our result clearly 
reproduces the asymmetry observed in Fig.\ 2 of Ref.\ \cite{SchTorBoy03}. 

The physical mechanism behind this asymmetry can be understood from the spatially resolved dynamics of the optical and matter-wave fields as described by Eqs.\ (\ref{sdeq}). For the sake of 
illustration, the spatial distribution of the condensate, the first-order 
atomic side-modes and the optical-field modes at the end of the strong pulse 
(i.e., at time $t_f$) are plotted in Fig.\ \ref{fig1}(b). 
Clearly, the atomic side-modes and the optical-field modes are well localized 
near the condensate edges. This preferential growth of the 
optical fields can be explained in the context of 
Eqs.\ (\ref{e_p}) and (\ref{e_m}) which imply 
that (at least for short times) the electric fields $\cE_\pm (z,t)$ grow 
monotonically in the $z$ and $-z$ directions, respectively, and are strongest 
at the ends of the condensate. As a result, according to Eq.\ (\ref{env_psi}), 
the scattering process is strongest in these areas of large electric fields, 
and the recoiling atoms mainly originate from the edges of the condensate. 
During the free time evolution following the strong pulse,
the forward-scattered atoms in the side modes $(1,\pm 1)$ initially travel towards the center of the BEC, whereas the atoms in the backwards side modes $(-1,\pm 1)$ immediately move away from the center [dashed lines in Fig.\ \ref{fig1}(a)]. The net effect is the observed 
asymmetry in the spatial distribution of forward and backward peaks.
So, in agreement with the experiments, our model predicts enhanced photon 
scattering near the edges of the condensate in the strong-pulse regime.  
This effect strongly supports an interpretation of the strong-pulse regime 
in the framework of {\em atomic scattering from optical fields} as was first 
conjectured by Ketterle and co-workers \cite{SchTorBoy03}.
\\
\begin{figure}
\centerline{\includegraphics[width=6.0cm]{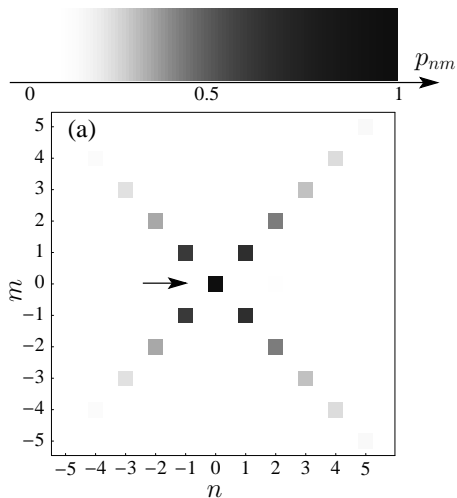}}
\caption{Spatially dependent model. 
  Atomic side-mode distributions in the strong-pulse regime for 
  $g=3.0\times 10^6\,{\rm s}^{-1}$ and $t_f=12\,\mu$s. The gray level of
  each square represents the ``relative" probability $p_{nm}$ as defined in the text. 
  The arrow indicates the direction of the incoming laser pulse and points towards the BEC ``side mode" $(0,0)$.
\label{XVsd}}
\end{figure}

\begin{figure}
\centerline{\includegraphics[width=7.cm]{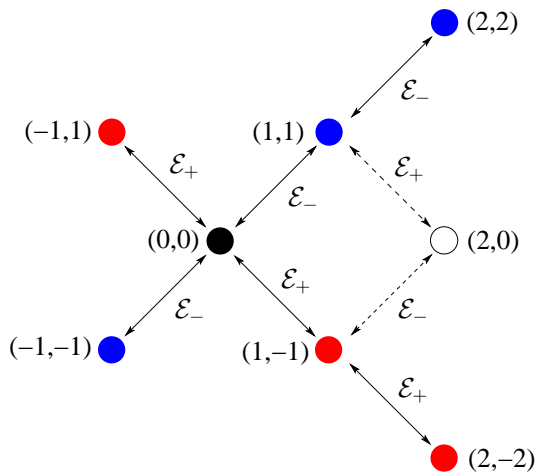}}
\caption{(Color online) Origin of the X-pattern: schematic representation.
Arrows show population transfer between different side modes $(n,m)$ 
via couplings to the optical fields ${\cal E}_\pm$. The side mode $(2,0)$ does not become populated due to the missing spatial overlap between the side modes $(1,\pm 1)$ and the endfire modes $\cE_{\pm}$, respectively.
\label{OriginX}}
\end{figure}

Our model also explains the experimentally observed X-shape patterns 
[see Fig.\ 1(A) of Ref.\ \cite{SchTorBoy03}]. 
In Fig.\ \ref{XVsd}, we show the atomic side-mode distribution 
forming an X-shape pattern at the end of a strong pulse with 
$g=3\times 10^6\, {\rm s}^{-1}$. The gray level of each square represents the ``relative" probability $p_{nm}=P_{nm}/P_{nm}^{\rm (max)}$ with $P_{nm} = \int dz |\psi_{nm}(z,t)|^2$ and $P_{nm}^{\rm (max)} = {\rm max}_{\{n,m\}}P_{nm}$. As a matter of fact, the enhanced 
photon scattering near the edges of the condensate is also 
responsible for the characteristic X-shape distribution of the side modes. 
Indeed, the appearance of an X-shape pattern requires the suppression 
of all off-diagonal side modes with $|n|\ne |m|$ and this is possible only if 
the atomic side modes and the optical endfire modes are located at 
the condensate edges. To make this point clear, let us consider, for example, 
the off-diagonal side mode $(2,0)$. Figure \ref{OriginX} shows the possible 
population transfers among the low-order atomic side-modes and the optical 
fields. According to Eq.\ (\ref{env_psi}), although the side mode 
$(2,0)$ is resonant 
with the modes $(1,\pm 1)$, it can only be populated if the $(1,\pm 1)$ modes 
overlap with the endfire modes $\cE_\pm$, respectively. 
As depicted in Fig.\ \ref{fig1}(b), however, this overlap is very small
since the modes are localized near the edges of the condensate, 
and the growth of the side mode $(2,0)$ is therefore suppressed. On the other hand, population transfer to the side modes $(2,\pm 2)$ is easily accomplished due to the strong overlap between the side modes $(1,\pm 1)$ and the fields $\cE_\mp$, respectively.

\subsection{Discussion}

Our studies show that the experimentally observed asymmetry and 
X-shape patterns are basically due to the spatial inhomogeneity of the 
optical fields. In an earlier attempt to theoretically describe these 
observations, Pu, Zhang, and Meystre \cite{PuZhaMey03} adopted a model which 
does not include any spatial propagation effects. In the framework of this 
model, they concluded that the observed asymmetry is due to angular effects, 
thus questioning the picture of atomic scattering from optical fields 
suggested by Ketterle and co-workers \cite{SchTorBoy03}. 
In particular, they found that forward recoiling atoms should have a symmetric 
distribution around $45^{\rm o}$, whereas backward recoiling atoms should favor 
larger angles due to a reduced energy mismatch. 

\begin{figure}
\centerline{\includegraphics[width=7.cm]{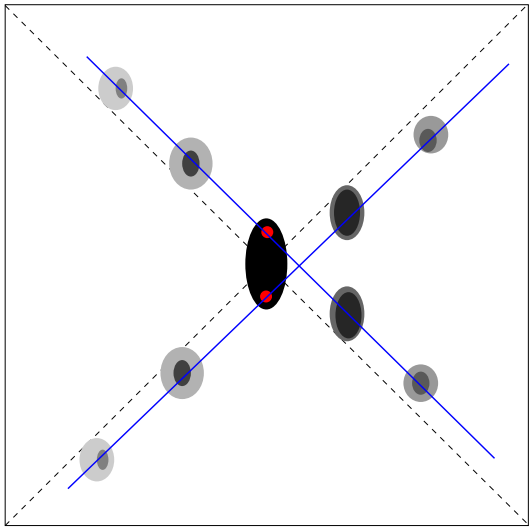}}
\caption{(Color online) Observed spatial asymmetry in an accurate schematic representation based on Fig.\ 1(A) of \cite{SchTorBoy03}.
The forward (backward) peaks clearly appear at angles smaller (larger) than 
$45^{\rm o}$. The dashed lines indicate the angles at $45^{\rm o}$.
\label{OriginAs}}
\end{figure}
 
Unfortunately, the analysis of \cite{PuZhaMey03} is restricted to short 
times for which the condensate remains practically undepleted and only 
first-order side modes are slightly populated. As a result there is no 
evidence that their theory is able to explain the characteristic 
X-shape pattern which involves higher-order atomic side modes. 
On the contrary, there are several different arguments which show that 
the explanation of \cite{PuZhaMey03} cannot account for the observed 
asymmetry and some of them have already been discussed elsewhere \cite{ZobNik05}. 
Here 
we would like to give a few more arguments which show the failure of the model 
employed in \cite{PuZhaMey03}. To this end, by analyzing Fig.\ 1(A) of 
\cite{SchTorBoy03} which depicts a typical experimental 
outcome, we have obtained a rather precise picture of the distribution 
of the atomic side modes which, for the sake of comparison, is presented   
in Fig. \ref{OriginAs}. 
Clearly, the atoms recoil at angles much smaller than $45^{\rm o}$ in the 
forward direction when measured with respect to the BEC center and the direction of the applied laser pulse (dashed lines in Fig.\ \ref{OriginAs}). The corresponding angle in the backward direction is larger than $45^{\rm o}$. 

These observations are incompatible with Fig.\ 3(b) of \cite{PuZhaMey03}. 
More precisely, this figure shows a very broad angular distribution 
for the forward and backward recoiling atoms. However, the angular width 
of the {\it individual} backwards and forwards travelling atomic 
wave packets observed in the experiments is actually much smaller. 
This means that the curves of Fig.\ 3(b) should be interpreted as 
probability distributions for the 
directions of recoiling atoms averaged over many realizations. 
In this case, however, Fig.\ 3(b)  of \cite{PuZhaMey03} 
shows that atoms should predominantly 
recoil at $45^{\rm o}$ in both forward and backward directions.  
Moreover, one should also observe {\em many} experimental runs where the 
forward recoiling atoms are emitted under a larger angle than the backwards 
recoiling ones. None of the experiments, however, reports on such types of 
observations. Furthermore, as depicted in our Fig.\ \ref{OriginAs}, experiments 
show the forward peaks to have a much smaller distance from the BEC center 
than the backward peaks. In an explication based on angular distributions, 
however, these distances have to be the same.  

These contradictions between theory and experiment clearly indicate that 
angular effects cannot account for the experimentally observed asymmetry 
and X-shape pattern. 
The main reason for the failure of the model employed 
in \cite{PuZhaMey03} is that it does not take spatial effects 
into account which, as we show here, are crucial in producing both 
the asymmetry and the X-shape pattern.  
Indeed, keeping only the two optical endfire modes, one can easily verify 
that in the semiclassical regime the model of \cite{PuZhaMey03} turns into 
a spatially-independent mean-field model similar to the ones employed 
e.g., in \cite{VasEfiTri04,RobPioBon04} and introduced at the end of Sec.\ II.

\begin{figure}[t]
\centerline{\includegraphics[width=6.0cm]{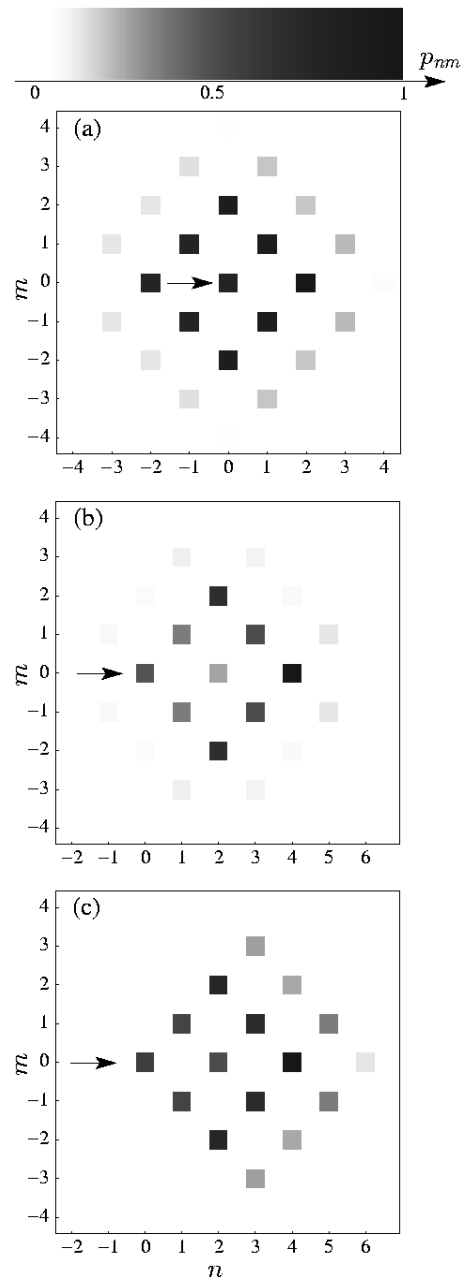}}
\caption{Spatially independent model. Atomic side-mode distributions (compare with Fig.\ \ref{XVsd}). 
  (a) Strong-pulse regime: $t_f=10.4\,\mu$s and $g=3.0\times 10^6\,{\rm s}^{-1}$;
  (b) weak-pulse regime:  $t_f=151.5\,\mu$s and $g=3.5\times 10^5\,{\rm s}^{-1}$;
  (c) weak-pulse regime:  $t_f=182\,\mu$s and $g=3.5\times 10^5\,{\rm s}^{-1}$. 
\label{XVsi}}
\end{figure}

As depicted in Fig.\ \ref{XVsi}(a), this simpler model cannot explain 
the characteristic X-shape pattern: There is no mechanism present to prevent the growth in off-diagonal side modes such as $(\pm 2,0)$ and $(0,\pm 2)$, and they soon become strongly 
populated. It is therefore obvious that spatial effects play a significant 
role in the strong-pulse regime.  

\begin{figure}[t]
\centerline{\includegraphics[width=7.cm]{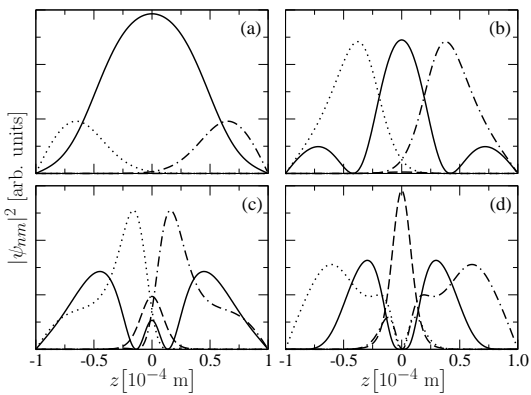}}
\caption{Spatially dependent model and weak-pulse regime
  ($g=3.0\times 10^5\,{\rm s}^{-1}$).
  Spatial distribution of the condensate (full curves), side modes $(1,1)$ (dotted), $(1,-1)$ (dash-dotted), and $(2,0)$ (dashed) at times   
  $t_f=200$ (a); $t_f=270$ (b); $t_f=350$ (c); 
  $t_f=420\,\mu$s (d).
\label{becWF_wp}}
\end{figure}

\section{Weak-pulse regime}

In this section, we will discuss some characteristic dynamic effects arising in the weak-pulse regime of BEC superradiance.
Together with Secs.\ III and IV, these results illustrate how our model allows to obtain new insights into the system behavior.
First of all, we describe typical stages of the spatially resolved time evolution of the matter-wave fields. In particular, we address the depletion of the condensate center reported in Ref.\ \cite{SchTorBoy03}.
Secondly, we discuss the time development of the optical fields and show that a minimum in the superradiant light emission does not imply that, at that time, the optical fields are also small {\it inside} the atomic sample. Finally, we show that our model is capable of reproducing the characteristic fan patterns for the side-mode distributions that have been observed experimentally. Additionally, we point out similarities and discrepancies in the predictions of the spatially dependent and independent models.

\subsection{\label{sec:mwf}Time evolution of matter-wave fields}

In this subsection, we examine characteristic features of the spatially resolved matter-wave dynamics in the weak-pulse regime. We describe the early stages in the time evolution of the $(1,\pm 1)$ and $(2,0)$ side modes -- which are resonantly coupled to the condensate -- and we discuss the depletion of the condensate center \cite{SchTorBoy03}.

A main difference between the weak- and strong-pulse regimes is the fact that in the latter only diagonal side modes with $|n|=|m|$ are populated significantly. As explained in Sec.\ IV, this is due to the missing spatial overlap between, e.g., the side modes $(1,\pm 1)$ and the endfire mode $e_{\pm}$ which prohibits the population of the $(2,0)$ and other non-diagonal side modes. Instead, higher-order diagonal side modes are populated, since this is not impeded by spatial effects and the strong fields allow to overcome the concomitant detuning barriers.

A different situation arises in the weak-pulse regime. In Figs.\ \ref{becWF_wp}, we show a typical scenario for the time evolution of the matter-wave fields. Behavior of this kind is observed for a large range of values of the external parameters. For very short times [Fig.\ \ref{becWF_wp}(a)], we have a situation similar to the strong-pulse regime, i.e., the first-order side modes $(1,\pm 1)$ start to grow at the edges of the condensate where the optical fields are strongest. The side mode $(2,0)$, although resonant with the modes $(1,\pm 1)$, cannot yet be populated because the necessary overlap between matter-wave and optical fields is missing. However, since the applied laser field is weak, higher-order diagonal side modes are now populated less efficiently because of the detuning barrier.

Instead, a sort of Rabi oscillation sets in between the first-order side modes and the condensate: after the condensate population at some point $z$ has been pumped completely to the first-order side mode, it is subsequently transferred back. This leads to the appearance of a minimum in the condensate density (marking the point of complete transfer) and a concomitant regrowth of the population from the edges [Fig.\ \ref{becWF_wp}(b)]. As to be expected, the minima stay near the density maxima of the first-order side modes. Once these maxima get close the center of the sample, the overlap between the $(1,\pm 1)$ side modes and the endfire mode $e_{\pm}$ becomes large, and the $(2,0)$ side mode starts to grow rapidly. Simultaneously, the minima in the condensate density merge at the center [Figs.\ \ref{becWF_wp}(c) and \ref{becWF_wp}(d)]. This scenario thus shows that there is a clear correlation between the onset of population growth in the $(2,0)$ side mode and the depletion of the condensate center. We expect that this effect should be observable experimentally. One should also note the localization of the $(2,0)$ side mode around the center of the system.

\begin{figure}[t]
\centerline{\includegraphics[width=7.cm]{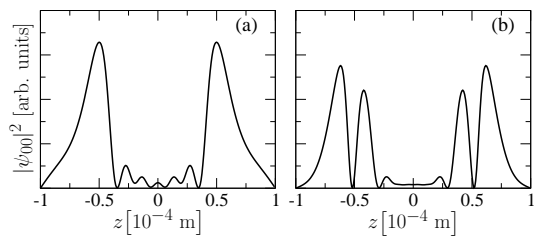}}
\caption{Depletion of condensate center in the weak-pulse regime
  ($g=6.0\times 10^5\,{\rm s}^{-1}$).
  Spatial distribution of the condensate at times
  $t_f=215$ (a) and $t_f=350\,\mu$s (b).
\label{becWF_depl}}
\end{figure}

In Ref.\ \cite{SchTorBoy03}, depletion of the condensate center was reported after the system had already undergone a large number of superradiant emission cycles and high-order side modes were populated. As evidenced in Figs.\ \ref{becWF_depl}, our simulations are able to reproduce this effect (note that $g$ is increased by a factor of 2 compared to Fig.\ \ref{becWF_wp}). At the times shown in these pictures, side modes of order $n=3$ and $4$ are populated, respectively. We find that the depletion of the center is clearly maintained over a long period of time. It should be noted, however, that the central population increases again briefly after the initial depletion described above. The simulations also show that the depletion effect becomes less pronounced if the pulse strength is reduced.

\subsection{Dynamics of optical fields}

\begin{figure}[t]
\centerline{\includegraphics[width=7.cm]{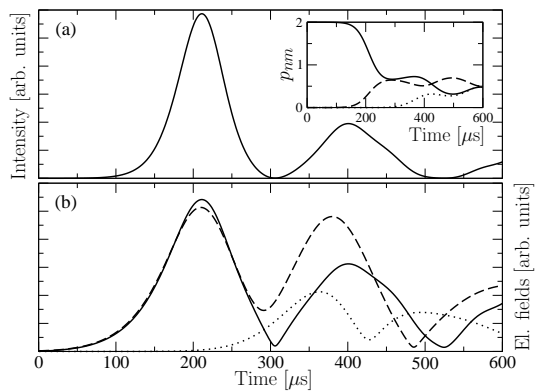}}
\caption{(a) Emitted superradiant light intensity as a function of time for $g=3.0\times 10^5\,{\rm s}^{-1}$. Inset: Populations of condensate (full curve), side modes $(1,\pm 1)$ (dashed) and $(2,0)$ (dotted). (b) Moduli of total electric field $|e_+|$ (full curve) and components $|e_+^{(0,0)}|$ (dashed) and $|e_+^{(1,1)}|$ (dotted).
\label{intensity}}
\end{figure}

Further insight into the characteristics of BEC superradiance can be obtained from studying the dynamics of the optical endfire modes. In Fig.\ \ref{intensity}(a) we show the emitted light intensity, which is proportional to the squared field strengths $|e_{\pm}|^2$ at the right and left edges of the condensate, respectively. The chosen coupling strength $g=3.0\times 10^5{\rm s}^{-1}$ is the same as in Fig.\ \ref{becWF_wp}, and the depicted behavior is again representative for the weak-pulse regime. For this value of $g$, only the side modes $(1,\pm 1)$ and $(2,0)$ become significantly populated in addition to the condensate [see inset in Fig.\ \ref{intensity}(a)]. To analyze the behavior of the intensity, we use Eq.\ (\ref{e_p}) to decompose, e.g., the endfire-mode field $e_+$ as $e_+(\xi,\tau) = \sum_{n,m} e_+^{(n,m)}(\xi,\tau)$ with
\be
e_+^{(n,m)}=\!-i\frac{\kappa}{\chi}\int_{-\infty}^\xi d\xi' e^{i(n -m)\tau}\non
 \psi_{nm}(\xi',\tau)\psi_{n+1,m-1}^*(\xi',\tau).
\ee
Each term $e_+^{(n,m)}$ can be interpreted as the contribution to the total field strength $e_+$ arising from transitions between the side modes $(n,m)$ and $(n+1,m-1)$. In Fig.\ \ref{intensity}(b), we plot the moduli $|e_+|$, $|e_+^{(0,0)}|$, and $|e_+^{(1,1)}|$. This diagram shows that the first maximum in the radiated intensity [cf.\ Fig.\ \ref{intensity}(a)] can be attributed to the transition between the condensate and the first-order side modes. The second maximum, however, is only partially caused by the higher-order transition to the side mode $(2,0)$, but has also a large component due to the continuing Rabi oscillations between the BEC and the side modes $(1,\pm 1)$ described in Sec.\ \ref{sec:mwf}. 

\begin{figure}[t]
\centerline{\includegraphics[width=7.cm]{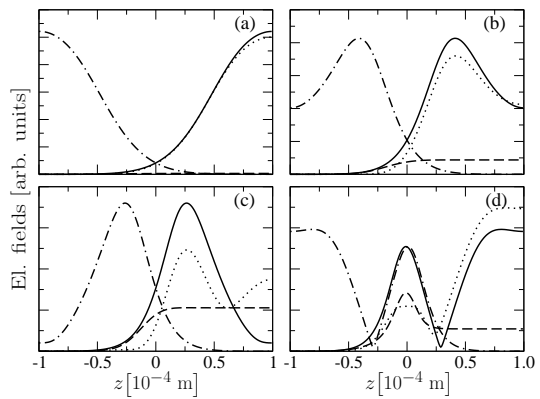}}
\caption{Spatial distribution of optical endfire modes for $g=3.0\times 10^5\,{\rm s}^{-1}$ and times $t_f=200$ (a); $t_f=270$ (b); $t_f=305$ (c); $t_f=420\,\mu$s (d). Full curve, $|e_+|$; dotted, $|e_+^{(0,0)}|$; dashed, $|e_+^{(1,1)}|$; dash-dotted, $|e_-|$.
\label{endfire}}
\end{figure}

In Figs.\ \ref{endfire}, we show the spatial dependence of the optical endfire modes at different times. For some of these times, the corresponding atomic side-mode distributions are displayed in Figs.\ \ref{becWF_wp}. Initially, the fields increase monotonically towards the edges of the condensate [Fig.\ \ref{endfire}(a), compare with Fig.\ \ref{becWF_wp}(a)]. After some time, however, the fields acquire a maximum {\it inside} the atomic sample [Fig.\ \ref{endfire}(b), see also Fig.\ \ref{becWF_wp}(b)]. This is due to the fact that near the edges, atoms are transferred back from the side modes $(1,\pm 1)$ into the condensate as mentioned in Sec.\ \ref{sec:mwf}. This requires the absorption of photons from the endfire modes and their subsequent emission into the laser field. As a consequence, the endfire-mode intensity is reduced. In Fig.\ \ref{endfire}(c), the field strength at the edges of the condensate, which determines the emitted superradiant light intensity, has decreased to almost zero. However, {\it inside} the atomic sample the fields are still strong. This means that a small emitted light intensity does not imply that the fields are also vanishing within the sample. Furthermore, from Fig.\ \ref{endfire}(c) we see that the field components $e_+^{(0,0)}$ and 
$e_+^{(1,1)}$ are still strong at the condensate edge, i.e., the vanishing of the total field strength can be considered as an interference effect. For later times, the behavior of the the fields and the various components becomes more complex as evidenced by Fig.\ \ref{endfire}(d) [corresponding to Fig.\ \ref{becWF_wp}(d)]. 
 
\subsection{Side-mode patterns}

\begin{figure}[t]
\centerline{\includegraphics[width=7.cm]{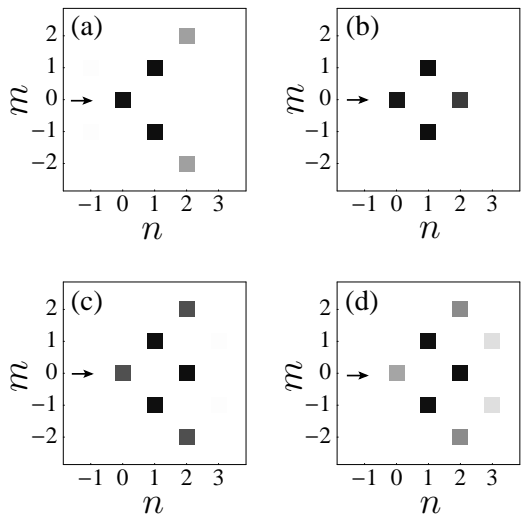}}
\caption{ 
Atomic side-mode distributions in the weak-pulse regime. The diagrams are for $g=4.5\times 10^5\,{\rm s}^{-1}$ and pulse durations  $t_f=140$ (a); $t_f=205$ (b); $t_f=260$ (c); $t_f=335\,\mu$s (d). Gray level as in Fig. \ref{XVsi}.
\label{SD_sidemodes_A}}
\end{figure}

\begin{figure}[t]
\centerline{\includegraphics[width=7.cm]{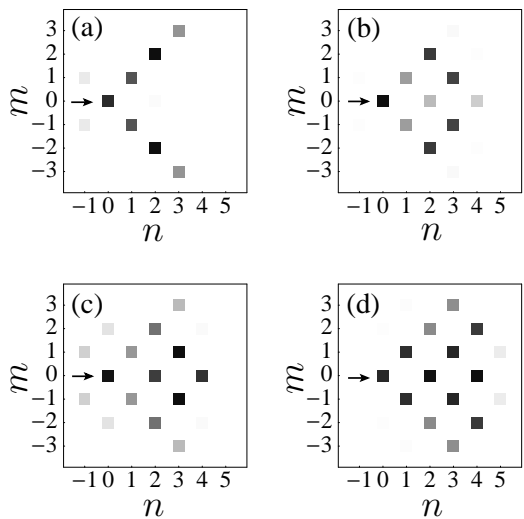}}
\caption{ 
As Fig.\ \protect\ref{SD_sidemodes_A}, but for $g=7.0\times 10^5\,{\rm s}^{-1}$ and pulse durations  $t_f=80$ (a); $t_f=170$ (b); $t_f=210$ (c); $t_f=280\,\mu$s (d). Gray level as in Fig. \ref{XVsi}.
\label{SD_sidemodes_B}}
\end{figure}

In the experiments on sequential superradiance in the weak-pulse regime, the atomic side-mode distributions were observed to display a characteristic fan shape [see, e.g., Figs.\ 1(F),(G) of Ref.\ \cite{InoChiSta99} and Fig.\ 1(B) of Ref.\ \cite{SchTorBoy03}]. As shown by Figs.\ \ref{SD_sidemodes_A} and \ref{SD_sidemodes_B}, our theoretical model is able to reproduce these patterns. Each of these figures displays atomic side mode distributions for a certain coupling strength and varying pulse durations. Clearly, the patterns shown in Figs.\ \ref{SD_sidemodes_A}(b)-\ref{SD_sidemodes_A}(d), \ref{SD_sidemodes_B}(b), and \ref{SD_sidemodes_B}(d) closely resemble the experimentally observed distributions mentioned above. Together with Figs.\ 2(b),(c) of Ref.\ \cite{ZobNik05}, these figures demonstrate that the emergence of such patterns is a typical outcome of our simulations over a large parameter range in the weak-pulse regime.

We also point out the similarity of our Fig.\ \ref{SD_sidemodes_B}(c) to Fig.\ 2(c) of Ref.\ \cite{YoSuToKu04}. In particular, both figures show population of the side modes$(0,\pm 2)$ which is not evident in Refs.\ \cite{InoChiSta99,SchTorBoy03}.
However, we also find a characteristic pattern which has not been reported experimentally so far and in which mainly side modes $(n,m)$ on the ``forward diagonals", i.e., for $|m|=n$, $n>0$, become populated [see Fig.\ \ref{SD_sidemodes_A}(a) and Fig.\ \ref{SD_sidemodes_B}(a)]. This pattern typically arises for short pulse durations before the full fan patterns develops. In this way, it appears to be intermediate between the X and the fan shape. It would be interesting to determine whether such patterns can also be observed experimentally.  
So far, however, the side mode distributions and their dependence on external parameters such as pulse strength and duration have not yet been studied systematically in experiments.

In contrast to the X shape patterns, fan-like side-mode patterns can also be obtained with the simpler mean-field model using spatial independent amplitudes [see Figs.\ \ref{XVsi}(b) and \ref{XVsi}(c)]. This implies that the role of spatial effects for pattern formation is somewhat less significant in the weak-pulse regime. Nevertheless, the fan shape is not only due to the varying detunings between the different side modes, but crucially depends on the coupled dynamics between the optical and matter-wave fields. We conclude this from studies of the mean-field model (\ref{sieq}) with the time-dependent optical endfire modes replaced by constants. In this case, the side-mode dynamics strongly deviates from the behavior found in the other models.

\begin{figure}[t]
\centerline{\includegraphics[width=7.cm]{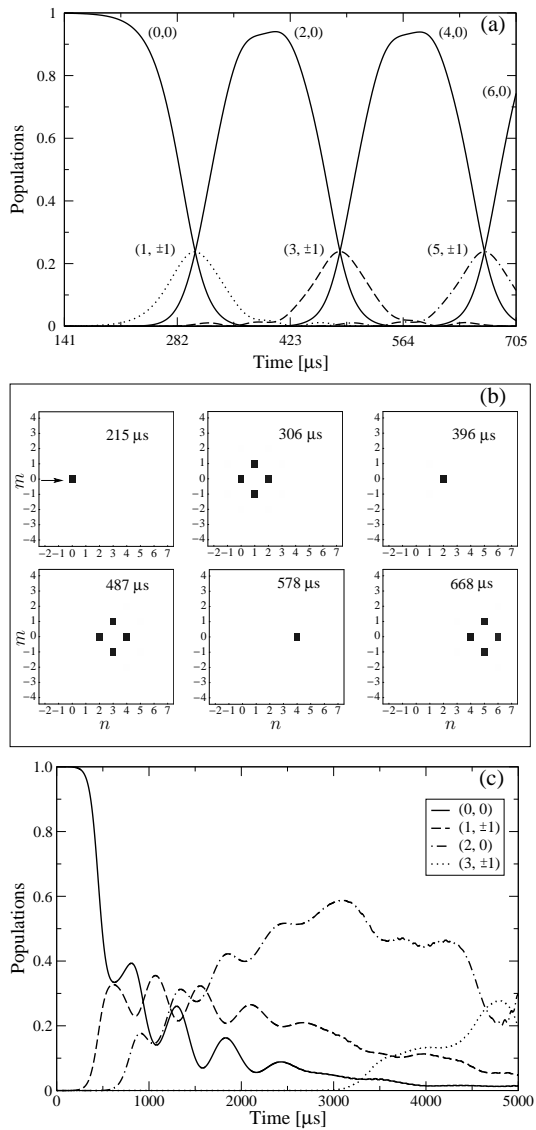}}
\caption{ Weak-pulse regime ($g=2.0\times 10^5\,{\rm s}^{-1}$).
Time evolution of side-mode populations (a) and distributions (b) for 
the spatially independent model (\protect\ref{sieq}). 
(c) Time evolution of side-mode populations for the spatially dependent model. 
Gray level as in Fig. \ref{XVsi}.
\label{SI_steady}}
\end{figure}

Although the spatially-independent mean-field model can 
reproduce fan-like patterns, significant differences to the spatially 
dependent model arise when studying very weak pulses and long times. 
These differences are illustrated in Fig.\ \ref{SI_steady} which shows 
the time evolution of the side mode populations in both models for a small 
value of the coupling constant. In the spatially independent model, 
we see that the population is successively transferred in a very regular 
way from the condensate to the side modes 
$(1,\pm 1)$, $(2,0)$, $(3,\pm 1)$, $(4,0)$, etc. 
[Figs.\ \ref{SI_steady}(a) and \ref{SI_steady}(b)]. 
Taking propagation effects into account, however, we see that the 
population remains trapped in the side modes  $(1,\pm 1)$ and $(2,0)$ 
for a very long time, and the regularity in the time evolution is 
apparently lost completely [see Fig. \ref{SI_steady}(c)].

\section{Conclusions}
We have presented a detailed theoretical analysis of superradiant Rayleigh 
scattering from BECs in the framework of the spatially dependent semiclassical 
Maxwell-Schr\"odinger equations. Our theoretical approach is rather general  
and equally well applies to many other problems involving interaction between 
light and ultracold atoms, e.g., matter-wave amplification \cite{InoPfaGup99,KozSuzTor99}. For the problem under consideration, the model allows us to reproduce and explain 
several characteristic features observed in the experiments that have not 
been theoretically accounted for so far.

For the strong-pulse regime our studies show that both the 
spatial asymmetry between backwards- and forwards-scattered atoms as well as the X-shape side mode patterns are consequences 
the spatial inhomogeneity of the optical fields. Such effects cannot be  
explained in the framework of spatially independent models. Our results strongly 
support an interpretation of the strong-pulse regime in terms of 
atomic scattering from optical fields, thus resolving an open controversy as explained in Sec.\ IV. 

In the weak-pulse regime, our model is able to reproduce the 
experimentally observed fan patterns of the side-mode distributions.  
Moreover, we have also obtained another kind of pattern which has not been experimentally reported so far and which can be regarded as intermediate to the X-shape strong-field pattern.
We have presented a detailed investigation of the spatially-resolved coupled dynamics of matter-wave and optical fields and pointed out characteristic effects. In particular, we have explained in detail that there is a clear correlation between the onset of population growth in the side mode $(2,0)$ and the depletion of the condensate center. In discussing the dynamics of the optical fields, we have shown that a minimum in the superradiant light emission does not necessarily imply that, at the same time, the optical fields are 
also small {\it inside} the atomic sample.

Finally, we have also given an analytical discussion of the short-time growth regime of the superradiant process. We have shown that the propagation effects turn the exponential growth expected from the spatially independent model into a slowed-down subexponential growth. 

Many features and effects that we have discussed throughout this work 
(e.g., the relation between the onset of population in the side mode $(2,0)$ and the 
depletion of the condensate center, or the appearance of intermediate side-mode 
patters) should be experimentally observable.  
We therefore hope that our results will stimulate further experimental 
investigations into the topic.    

There are several directions for further theoretical work on the basis of the results presented here. First of all, it is of interest to incorporate the effects of the initial quantum-mechanical noise, that starts up the process, more fully into the theory and to discuss the ensuing fluctuations in the macroscopic system dynamics. First steps in this direction were reported in \cite{ZobNik05}. Secondly, as mentioned above, the present theory can directly be applied to study matter-wave amplification. In this way it should be possible to obtain an improved control over the process. Finally, it would also be important to extend the present one-dimensional treatment of the superradiant dynamics to a full three-dimensional theory in order to study transverse propagation effects more closely, such as multimode radiation.

\appendix*
\section{}

We introduce the functions $Q(\xi,\tau) = \psi_{1,-1}(\xi,\tau)$, $B(\xi,\tau) = \psi_{-1,1}^*(\xi,\tau)e^{2i\tau}$, and $E(\xi,\tau) = \int_{-\infty}^\xi d\xi' [Q(\xi',\tau)+B(\xi',\tau)]$. Using $\psi_{00}(\xi,\tau) = \sqrt{N/\Lambda}$ and setting $\mGamma = \Gamma/\Lambda$, Eqs.\ (\ref{startup}) can be written as
\begin{subequations}\label{app1}
\bea
\frac{\partial E(\xi,\tau)}{\partial \xi} &=& Q(\xi,\tau)+B(\xi,\tau), \\
\frac{\partial Q(\xi,\tau)}{\partial \tau} &=& \mGamma E(\xi,\tau),\\
\frac{\partial B(\xi,\tau)}{\partial \tau} &=& -\mGamma E(\xi,\tau)+2iB(\xi,\tau).
\eea
\end{subequations}
We wish to solve Eqs.\ (\ref{app1}) under the boundary conditions $Q(\xi,0)=\psi_{1,-1}(\xi,0)$, $B(\xi,0) = 0$, and $E(0,\tau)=0$. Our starting-point is the Laplace transform method applied in Ref.\ \cite{RayMos81} to a similar set of equations. We thus define the transforms $e(s,\tau) = {\cal L}\{E(\xi,\tau)\}\equiv \int_0^\infty e^{-s\xi} E(\xi,\tau)d\xi$, $q(s,\tau) = {\cal L}\{Q(\xi,\tau)\}$, and $b(s,\tau) = {\cal L}\{B(\xi,\tau)\}$. Since $E(0,\tau)=0$, Eqs.\ (\ref{app1}) transform to
\begin{subequations}\label{laplace}
\bea
se(s,\tau) & =& q(s,\tau) + b(s,\tau),\\
\label{lap2}\frac{\partial q(s,\tau)}{\partial \tau} &=& \mGamma e(s,\tau),\\
\label{lap3}
\frac{\partial b(s,\tau)}{\partial \tau} &=& -\mGamma e(s,\tau)+2ib(s,\tau).
\eea
\end{subequations}
After replacing $e(s,\tau)$ by $(q+b)/s$, Eqs.\ (\ref{lap2}) and (\ref{lap3}) turn out to have the same formal structure as the equations of motion for $C_{1,-1}(t)$ and $\widetilde C_{-1,1}(t)$ in the spatially independent model, and we can immediately deduce their solution from Eqs.\ (\ref{cpm}) and (\ref{cmp}) by replacing $\Gamma$ with $2\mGamma/s$. This yields
\bea\label{lapq}
q(s,\tau) &=& \frac{q(s,0)}{2[i-\lambda_1(s)]}\left[e^{\lambda_1(s) \tau}\left(\lambda_2(s) -\frac\mGamma s\right) \right. \non
&& \left. -e^{\lambda_2(s) \tau}\left(\lambda_1(s)-\frac\mGamma s\right)\right],\\ \label{lapb}
b(s,\tau) &=& \frac{q(s,0)}{2[i-\lambda_1(s)]} \frac\mGamma s \left(e^{\lambda_1(s) \tau} - e^{\lambda_2(s) \tau}\right)
\eea
with $\lambda_1(s) = i-\sqrt{-1-2i\mGamma/s}$ and $\lambda_2(s) = i+\sqrt{-1-2i\mGamma/s}$. In contrast to the case of Ref.\ \cite{RayMos81}, the inverse transform can no longer be carried out in closed form. In view of Eq.\ (\ref{decom}), we thus specialize $q(s,0) = 1/(s-2\pi i n/\Lambda)$ which is the Laplace transform of the exponential factors. The inversion integral
\be\label{bromwich}
Q(\xi,\tau ) = \frac 1{2\pi i}\int_{x-i\infty}^{x+i\infty} ds\, e^{s\xi}q(s,\tau),
\ee
$x > 0$, is then evaluated by means of the saddle-point method (see, e.g., Ref.\ \cite{ArfWeb01}). Except for very short times, the terms in Eqs.\ (\ref{lapq}) and (\ref{lapb}) involving $\lambda_1$ can be neglected, since they only yield an exponentially decaying contribution. The relevant saddle point $s_0$ for the evaluation of Eq.\ (\ref{bromwich}) is then given by the solution of $\xi s_0 +\lambda_2(s_0)\tau = 0$. In general, $s_0$ has to be expressed as the root of a fourth-order polynomial. In the weak- and strong-pulse limits of $\Gamma \ll 1$ and $\Gamma\gg 1$, however, one can approximate $s_0 \approx \sqrt{\mGamma\tau}$ and $s_0 \approx e^{-i\pi/6} (\mGamma\tau^2/2\xi^2)^{1/3}$, respectively. Inserting these values into the saddle-point formula (and also approximating the accompanying slowly varying function appropriately), we obtain relations (\ref{psi_weak}), (\ref{psi_strong}), and (\ref{k_shape}). We have verified that the saddle-point method without these further approximations as well as -- within their respective ranges of validity -- the simplified formulas (\ref{psi_weak}), (\ref{psi_strong}), and (\ref{k_shape}) give very good approximations to the numerical solution of Eqs.\ (\ref{startup}). The result (\ref{psi_weak}) is consistent with the fact the solution of Eqs.\ (\ref{startup}) without the backward mode $\psi_{-1,1}(\xi,\tau)$ is given by $\psi_{1,-1}(\xi,\tau) = \psi_{0} I_0(2\sqrt{\mGamma\tau \xi})$ with the modified Bessel function $I_0(x)\approx \exp(x)/\sqrt{2\pi x}$, $x\gg 1$ (see also \cite{GroHar82}).

\end{document}